\documentclass[a4paper,twocolumn,showpacs,superscriptaddress,floatfix,prl,nofootinbib]{emulateapj}
\usepackage{bm}
\usepackage{multirow}
\usepackage{graphicx}
\usepackage{epsf}
\usepackage{epsfig}
\usepackage{amssymb,amsmath}
\usepackage[usenames]{color}
\usepackage{amssymb}
\usepackage{times}
\usepackage{hyperref}
\newcommand{\cf}{cf.~}
\newcommand{\ie}{i.e.~}
\newcommand{\eg}{e.g.~}

\begin{document}

\title{On the detectability of dual jets from binary black holes}

\author{Philipp Moesta\altaffilmark{1,4}, Daniela Alic\altaffilmark{1}, Luciano Rezzolla\altaffilmark{1,2}, Olindo Zanotti\altaffilmark{1}, Carlos Palenzuela\altaffilmark{3,2}}

\altaffiltext{1}{
  Max-Planck-Institut f\"ur Gravitationsphysik,
  Albert-Einstein-Institut,
  Potsdam-Golm, Germany
}
\altaffiltext{2}{
  Department of Physics and Astronomy,
 Louisiana State University,
  Baton Rouge, LA, USA
}
\altaffiltext{3}{
Canadian Institute for Theoretical Astrophysics,
 Toronto, Ontario M5S 3H8, Canada
}
\altaffiltext{4}{
  TAPIR, MC 350-17, 
  California Institute of Technology,
  Pasadena, CA, 91125 USA}


\begin{abstract}
We revisit the suggestion that dual jets can be produced during the
inspiral and merger of supermassive black holes when these are
immersed in a force-free plasma threaded by a uniform magnetic
field. By performing independent calculations of the late inspiral and
merger, and by computing the electromagnetic emission in a way which
is consistent with estimates using the Poynting flux, we show that a
dual-jet structure is present but energetically subdominant with
respect to a non-collimated and predominantly quadrupolar emission,
which is similar to the one computed when the binary is in
electrovacuum. While our findings set serious restrictions on the
detectability of dual jets from coalescing binaries, they also
increase the chances of detecting an EM counterpart from these
systems.
\end{abstract}

\maketitle

\section{Introduction}The inspiral and merger of supermassive
black holes (BHs) will represent a secure source for the planned
space-borne gravitational-wave (GW) detectors. Together with the GW
signal, this process is expected to be accompanied either
before~\cite{Bode2011, Farris2011} or after~\cite{Rossi2010,
  Zanotti2010} the merger by the emission of electromagnetic (EM)
radiation, thus providing a perfect example of multi-messenger
astronomy. Should a ``simultaneous'' detection take place, this would
not only help to localize the GW source and provide its redshift, but
also address a number questions in astrophysics and
cosmology~\cite{Haiman2009b,Sesana2011}.

As the merger between two galaxies takes place and the BHs get closer,
a circumbinary accretion disk is expected to form. Because the
radiation-reaction timescale over which the binary evolves is much
longer than the accretion timescale, the disk will slowly follow the
binary as its orbit shrinks. However, as the binary separation becomes
$\sim 10^5\,M$, where $M$ is the mass of the binary, the
radiation-reaction timescale reduces considerably, becoming smaller
than the accretion one. When this happens, the disk evolution
disconnects from the binary, the accretion rate reduces, and the
binary performs its final orbits in an inner region poor of
gas~\cite{Armitage:2002, Liu2003}.

This basic picture represents the astrophysical backdrop of a simple
model which has been used by a number of authors to model the EM
emission from the BH binary. More specifically, assuming that the disk
is threaded by a coherent and large-scale magnetic field which is
anchored to the disk, this will also permeate the ``evacuated'' region
where the binary is rapidly shrinking and provide an effective way of
coupling the binary's orbital motion to the generation of an EM
signal. This scenario has been considered both for spacetimes in
electrovacuum (EV)~\cite{Palenzuela:2009yr, Palenzuela:2009hx,
  Moesta:2009} and for spacetimes filled by a tenuous plasma in the
force-free (FF) approximation~\cite{Palenzuela:2010a,
  Palenzuela:2010b, Neilsen:2010}. Reference~\cite{Moesta:2009}, in
particular, has considered a series of spinning BH binaries and
studied the dependence of the gravitational and EM signals with
different spin configurations. All in all, it was found that EM
radiation in the lowest $\ell=2, m=2$ multipole accurately reflects
the gravitational one. Furthermore, the efficiency of the energy
emission in EM waves was found to scale quadratically with the total
spin and to be $\sim 13$ orders of magnitude smaller than the one in
GWs for realistic magnetic fields. However, the prospects of detecting
an EM counterpart have become larger when it was pointed out
in~\cite{Palenzuela:2010a} that during the inspiral in an FF plasma, a
\textit{dual-jet} structure forms, as a generalization of the
Blandford-Znajek process~\cite{Blandford1977} to orbiting
BHs.
Under these conditions, the EM energy can accelerate electrons
and lead to synchrotron radiation.

We extend here the investigations made in~\cite{Moesta:2009} by
considering the evolution in an FF regime and properly computing the EM
emission as the net result of ingoing and outgoing radiation. We
confirm in this way the presence of a dual-jet structure, but also
find that this is energetically subdominant with respect to a
non-collimated and predominantly \textit{quadrupolar} emission, at
least during the late inspiral and merger we have considered. Hence,
if the simplified scenario considered here is realized in
astrophysical configurations, it will be difficult to reveal the
emission from the dual jets, but it will also be easier that an EM
counterpart to binary BH mergers can be detected.

\section{Methodology}As mentioned above, we assume the
magnetic field to be anchored to the disk, whose inner edge is at a
distance of $\sim 10^3\, M$ and is effectively outside of our
computational domain, while the initial binary separation is only $D=
8\,M$. Although the large-scale magnetic field is poloidal, it is set
to be initially uniform within the computational domain, \ie $B^i =
(0,0,B_0)$ with $B_0 \sim 10^{-4}/M$. We note that although
astrophysically large, the initial magnetic field has an EM energy
several orders of magnitude smaller than the gravitational-field
one, so that the EM fields can be treated as test-fields and all
the results for the luminosity scale quadratically with $B_0$. The
electric field $E^i$ is initially zero, but it rapidly reaches a
consistent solution~\cite{Moesta:2009}. As BH initial data we consider
binaries with equal masses but with two different spin configurations:
namely, the $s_0$-binary (both BHs are nonspinning) and the
$s_6$-binary (both BHs have spins aligned with the orbital angular
momentum with $(J/M^2)_1 = a_1 = a_2 = 0.6$;
see~\cite{Rezzolla-etal-2007, Reisswig:2009vc} for details). These two
configurations allow us to study both the contributions coming from
the binary's orbital motion, but also those related to the spins of
the two BHs. The setup for the numerical grids used in the simulations
consists of nine levels of mesh refinement with a fine resolution of
$\Delta x/M=0.025$. The wave-zone grid has a resolution of $\Delta
x/M=1.6$ and extends from $r=24\,M$ to $r=180\,M$, while the outer
(coarsest) grid extends to $819.2\,M$. In our implementation of the FF
equations we solve the same set of equations
in~\cite{Palenzuela:2010b} but do not enforce ``by-hand'' the FF
condition. Rather, we implement a damping scheme which drives the
solution to satisfy the FF condition and avoids inconsistent EM
fields; we will present our approach in~\cite{Alic:2011b}.

The calculation of the EM and gravitational radiation generated during
the inspiral, merger and ringdown is arguably the most important
aspect of this work as it allows us to establish what are the
characteristics of the emissions in two channels. For the GW sector,
we compute the emission via the Newman-Penrose (NP) curvature scalar
$\Psi_4$ as detailed in~\cite{Pollney:2007ss:shortal} and~\cite{Moesta:2009}. 
In practice, we define an orthonormal basis in the
three space $(\hat{\boldsymbol{r}}, \hat{\boldsymbol{\theta}},
\hat{\boldsymbol{\phi}})$, with poles along
$\hat{\boldsymbol{z}}$. Using the normal to the slice as timelike
vector $\hat{\boldsymbol{t}}$, we construct the null orthonormal
tetrad $\{\boldsymbol{l}, \boldsymbol{n}, \boldsymbol{m},
\overline{\boldsymbol{m}}\}$, with the bar indicating a complex
conjugate
\begin{equation}
   \boldsymbol{l} = \frac{1}{\sqrt{2}}(\hat{\boldsymbol{t}} + \hat{\boldsymbol{r}}),\quad
   \boldsymbol{n} = \frac{1}{\sqrt{2}}(\hat{\boldsymbol{t}} - \hat{\boldsymbol{r}}),\quad
   \boldsymbol{m} = \frac{1}{\sqrt{2}}(\hat{\boldsymbol{\theta}} + i\hat{\boldsymbol{\phi}}) \,,
\end{equation}
by means of which we project the Weyl curvature tensor $C_{\alpha
  \beta \gamma \delta}$ to obtain $\Psi_4 = C_{\alpha \beta \gamma
  \delta} n^\alpha {\bar m}^\beta n^\gamma {\bar m}^\delta$. For the
EM sector, instead, we use two equivalent approaches to cross-validate
our measures. The first one uses the NP scalars $\Phi_0$
(for the ingoing EM radiation) and $\Phi_2$ (for the outgoing EM
radiation), defined using the same tetrad~\cite{Teukolsky73}
\begin{equation}
\Phi_0 \equiv F^{\mu\nu} l_{\nu} m_{\mu}\,, \qquad
\Phi_2 \equiv F^{\mu\nu} \overline m_{\mu} n_{\nu}\,.
\end{equation}
It is always useful to remark that by construction quantities such as
$\Psi_4, \Phi_0, \Phi_2$ are well-defined only at very large distances
from the sources. Any measure of these quantities in the
strong-field region risks to be incorrect. As an example, the EM
energy flux does not show the expected $1/r^2$ scaling when $\Phi_0,
\Phi_2$ are measured at distances of $r \simeq
20\,M$~\cite{Palenzuela:2010a, Palenzuela:2010b}, which is instead
reached only for $r \gtrsim 100\,M$. As we will show later, this
results in a considerable difference in the estimate of the
non-collimated EM emission.

Since our choice of having a uniform magnetic field within the
computational domain has a number of drawbacks (\eg nonzero initial
$\Phi_2\,, \Phi_0$), great care has to be taken when measuring the EM
radiation. Fortunately, we can exploit the linearity in the Maxwell
equations to distinguish the genuine emission induced by the presence
of the BH(s) from the background one. Following~\cite{Teukolsky73}, we
compute the total EM luminosity as a surface integral across a
2-sphere at a large distance
\begin{equation}
L_{_{\rm EM}} = \lim_{r \rightarrow \infty}  \int  
r^2 \left(|\Phi_2 - \Phi_{2,{\rm B}}|^2 - |\Phi_0 - \Phi_{0,{\rm
    B}}|^2\right)d\Omega\,,
\label{FEM_JLmt0}
\end{equation}
where $\Phi_{2,{\rm B}}$ and $\Phi_{0,{\rm B}}$ are the values of the
background scalars induced by the asymptotically uniform magnetic
field solution in the time dependent spacetime produced by the binary
BHs. The choice of $\Phi_{2,{\rm B}}$ and $\Phi_{0,{\rm B}}$ will
be crucial to measure correctly the radiative EM emission. 

The background values of the NP scalars $\Phi_{2,{\rm B}},
\Phi_{0,{\rm B}}$ to be used in Equation~\eqref{FEM_JLmt0} are themselves time
dependent and cannot be distinguished, at least a priori, from the
purely radiative contributions. A first prescription takes the
background values to be time independent and those at the initial time
\begin{equation}
 \Phi_{2,{\rm B}} = \Phi_{2} (t=0)\,, \qquad
 \Phi_{0,{\rm B}} = \Phi_{0} (t=0)\,.
\label{first_guess}
\end{equation}
A second prescription involves instead the removal of those multipole
components of the NP scalars which are not radiative, namely all those
associated with the $m=0$ multipoles
\begin{equation}
 \Phi_{2,{\rm B}} = (\Phi_{2})_{\ell, m=0}\,, \qquad
 \Phi_{0,{\rm B}} = (\Phi_{0})_{\ell, m=0}\,,
\label{second_guess}
\end{equation}
where $(\Phi_{2,0})_{\ell, m=0}$ refer to the $m=0$ modes of the
multipolar decomposition of $\Phi_{2,0}$ ($\ell \leq 8$ is sufficient
to capture most of the background). Note that because the $m=0$
background is essentially time independent (after the initial
transient), the choice~\eqref{second_guess} is effectively equivalent
to considering as background the final values of the NP scalars as
computed in an EV calculation of the same binary system.  

\begin{figure*}
  \begin{center}
     \vglue -0.7cm
     \hglue -0.7cm
     \includegraphics[angle=-90,width=6.0cm]{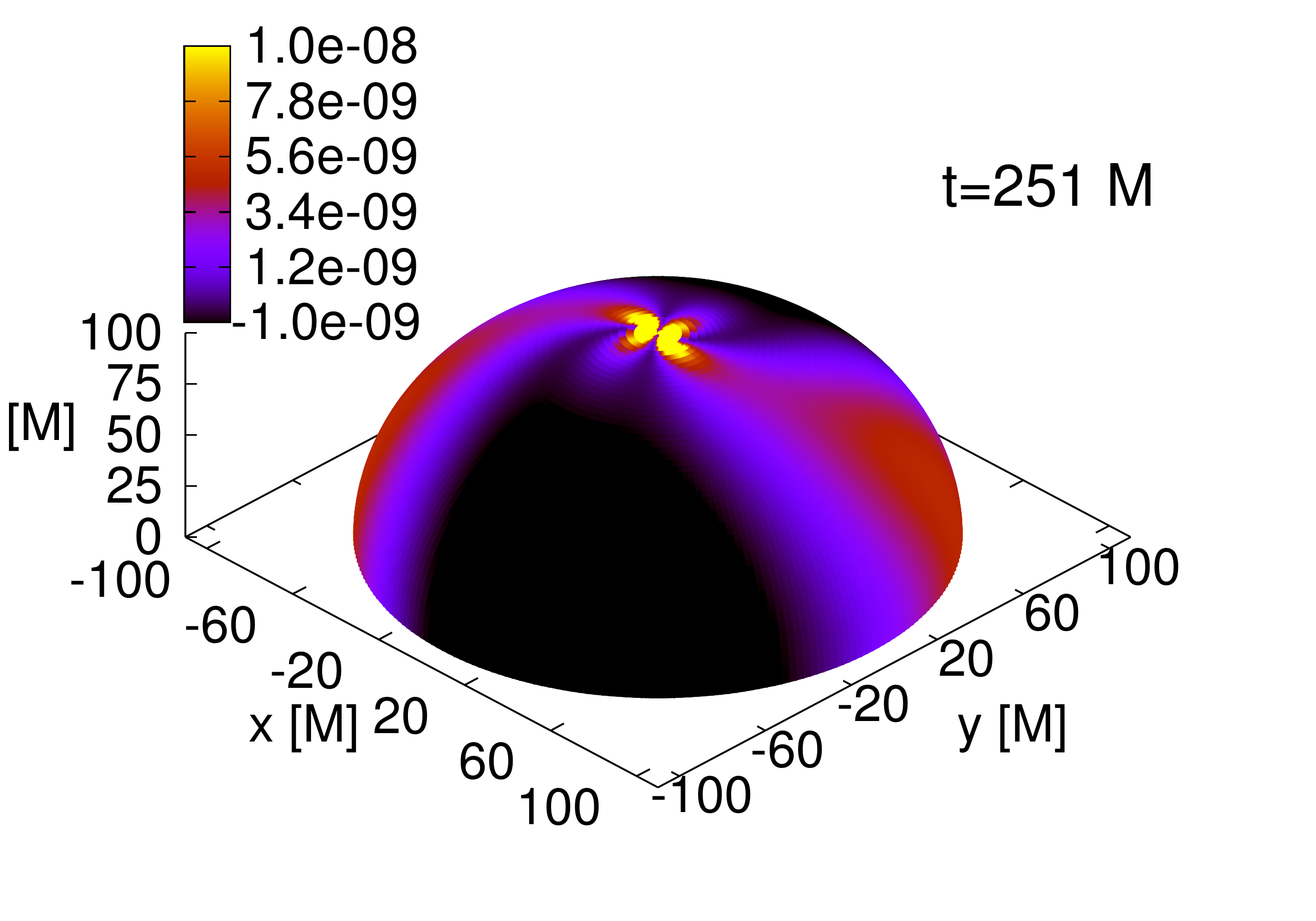}
     \hglue -0.7cm
     \includegraphics[angle=-90,width=6.0cm]{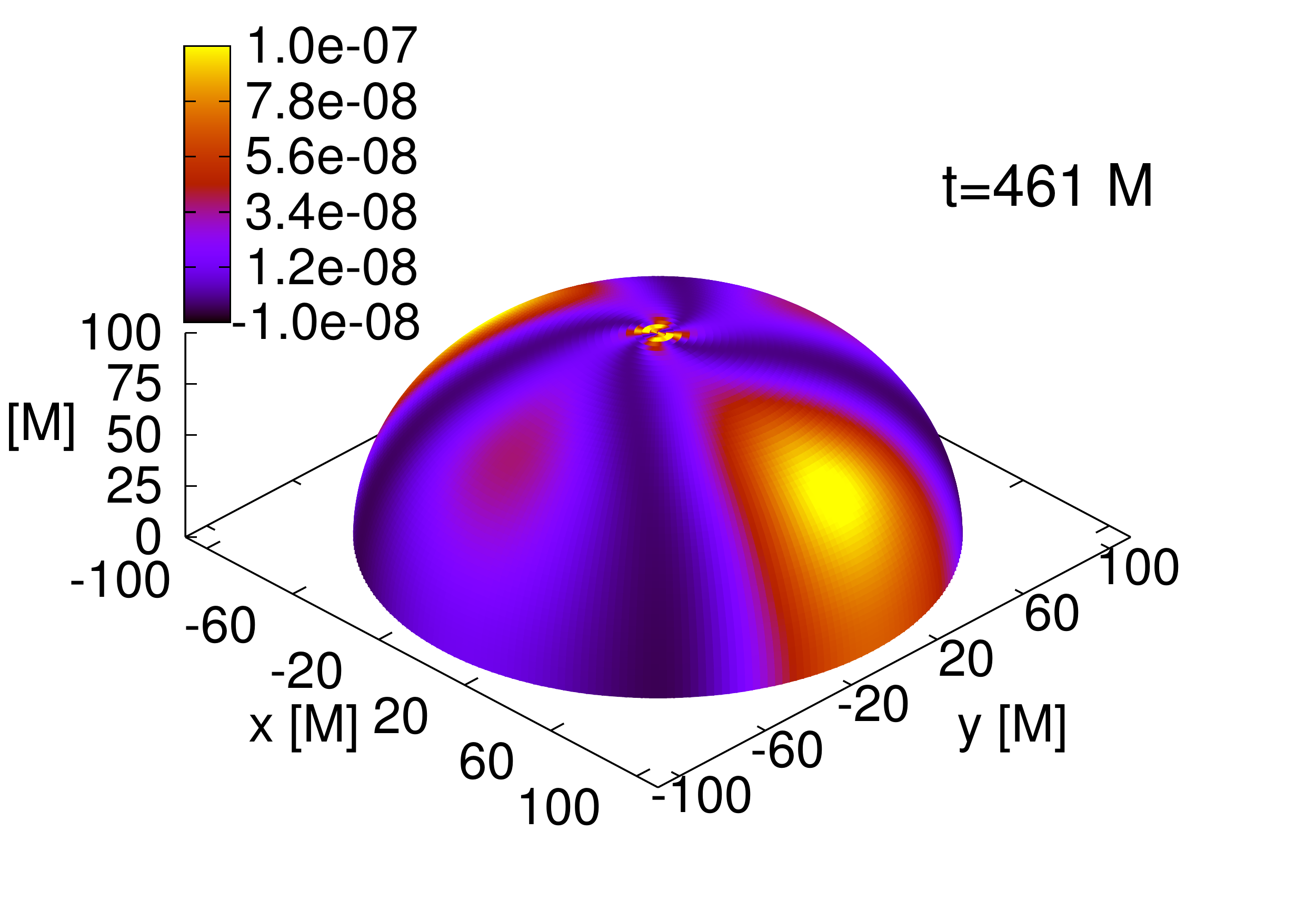}
     \hglue -0.7cm
     \includegraphics[angle=-90,width=6.0cm]{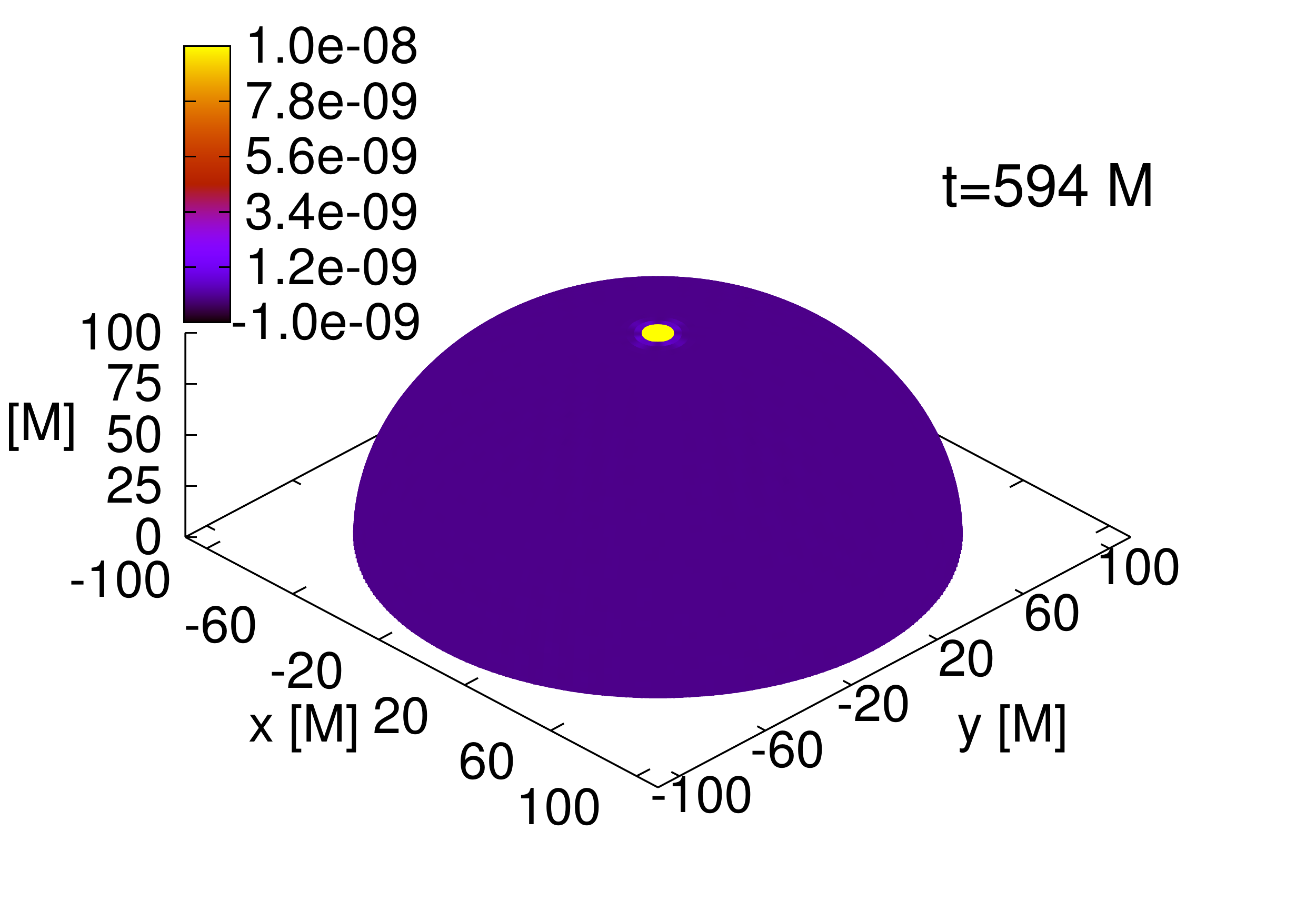}
     \vglue 0.3cm
     \hglue -0.7cm
     \includegraphics[angle=-90,width=6.0cm]{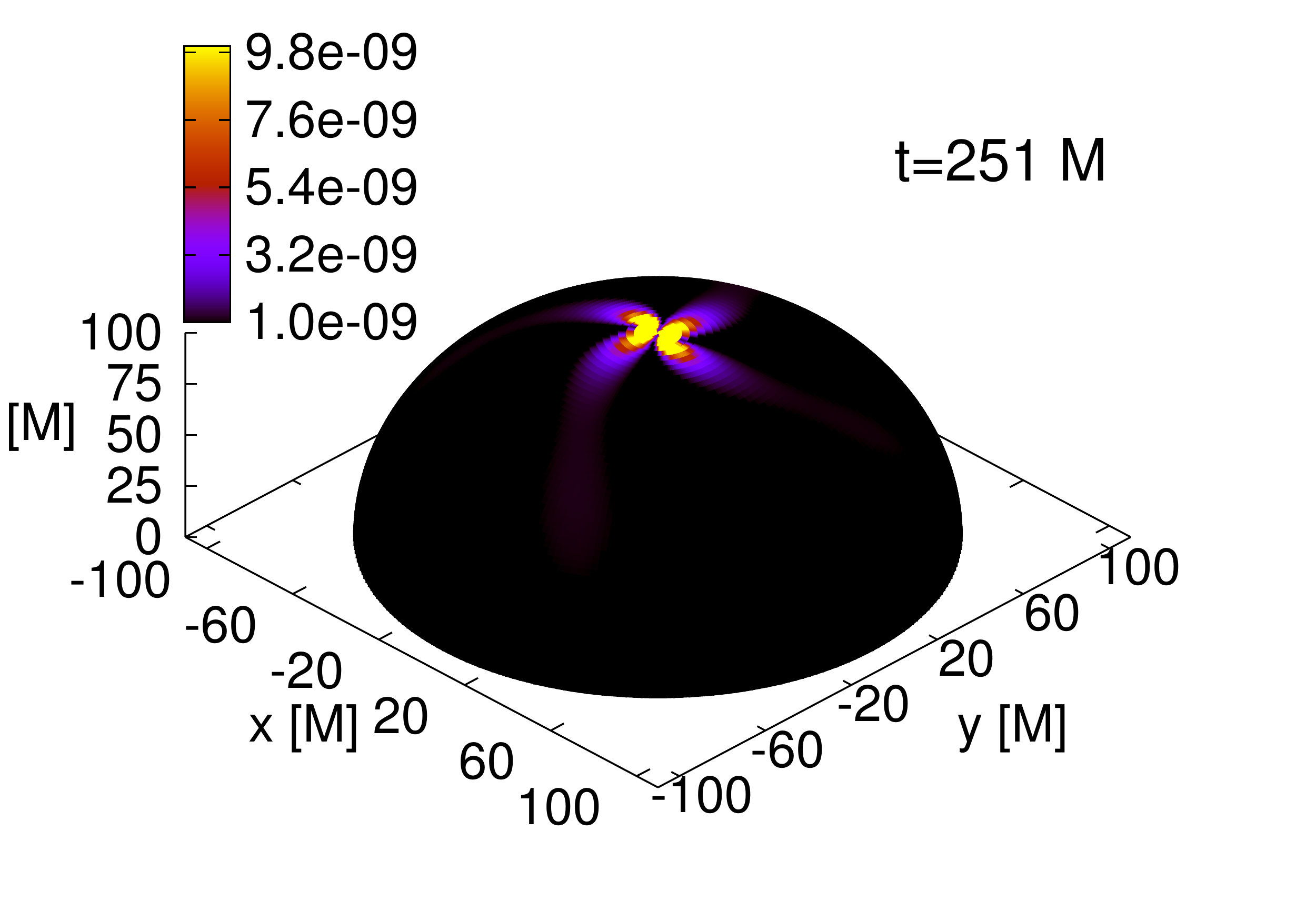}
     \hglue -0.7cm
     \includegraphics[angle=-90,width=6.0cm]{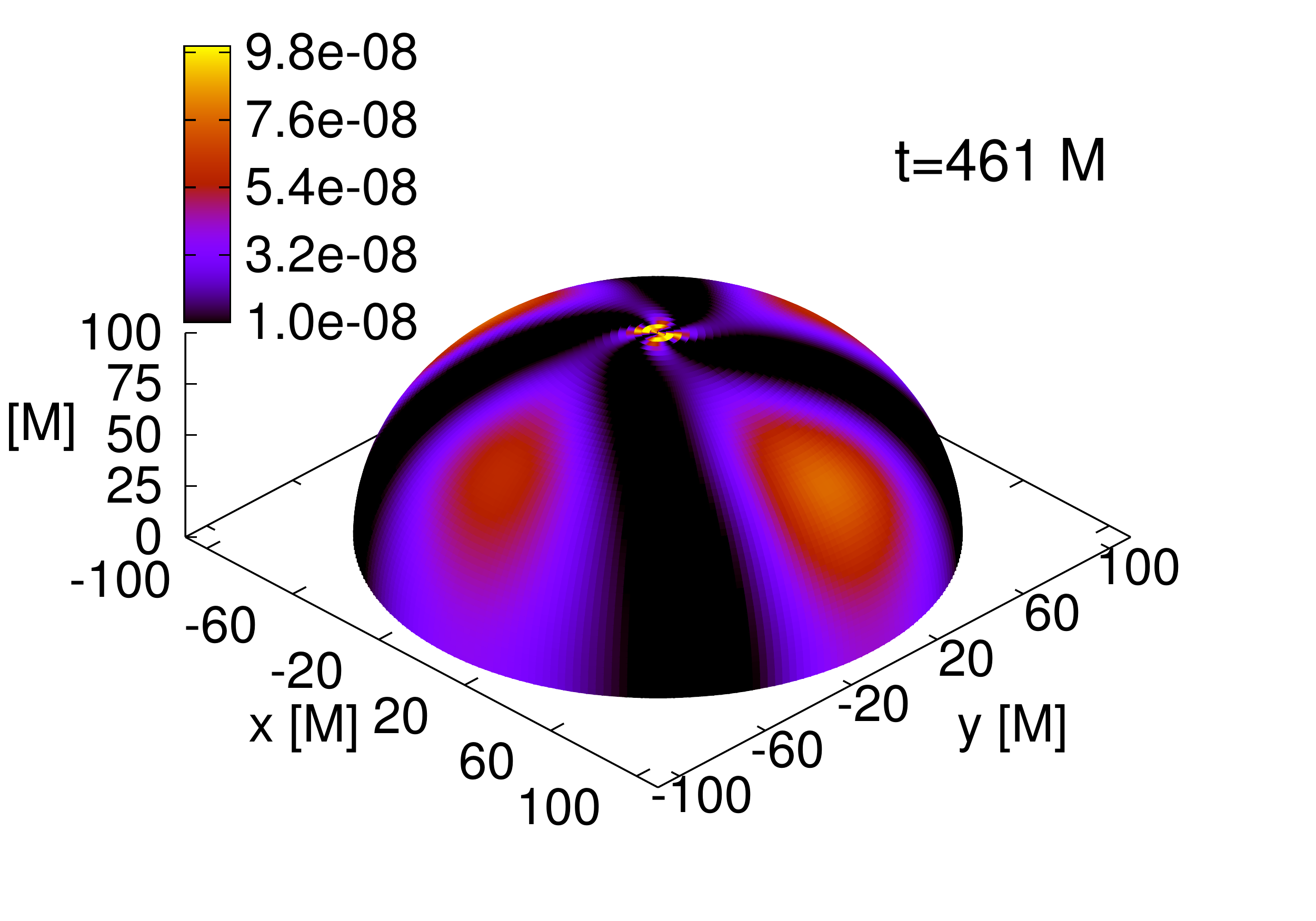}
     \hglue -0.7cm
     \includegraphics[angle=-90,width=6.0cm]{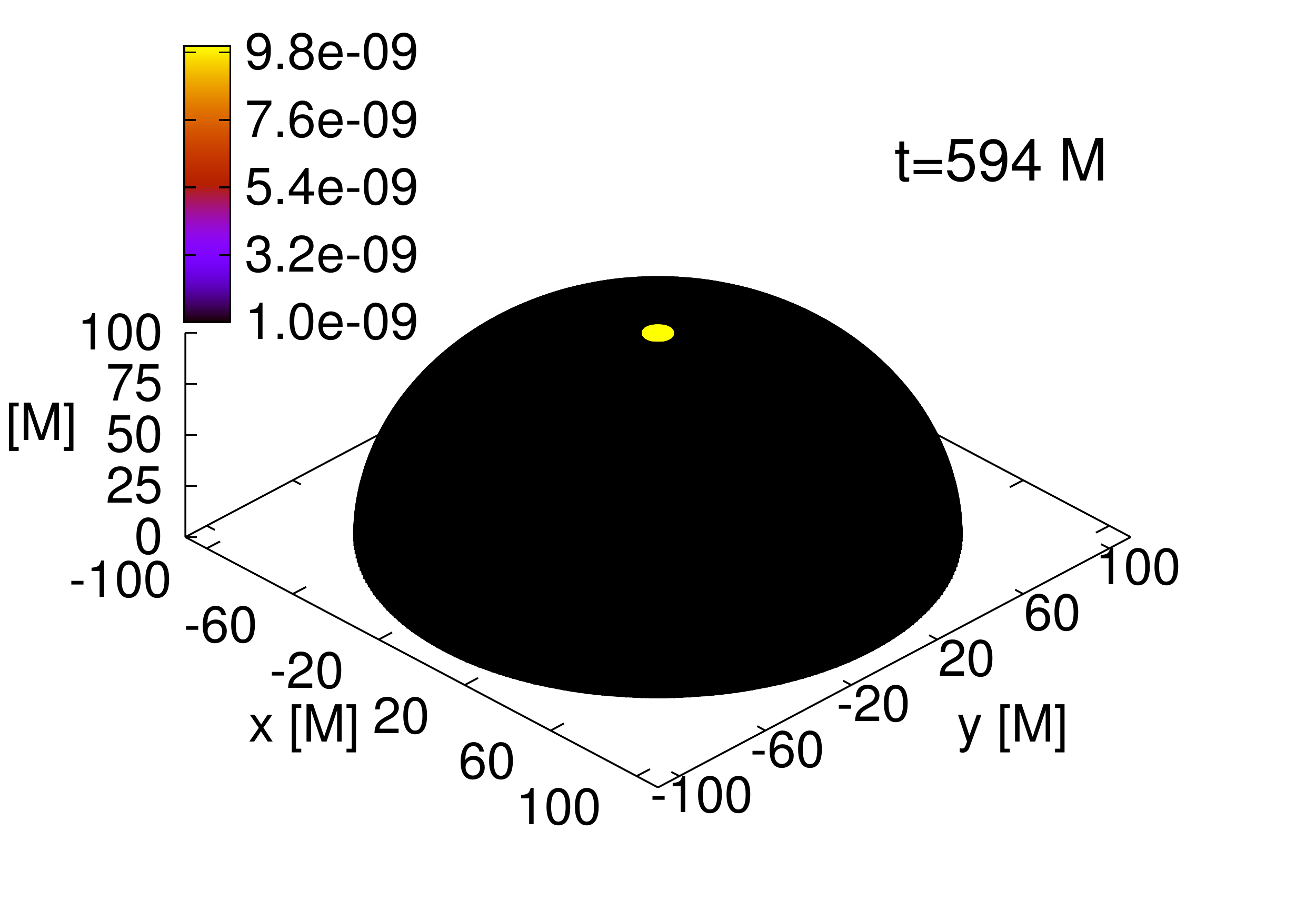}
  \end{center} 
    \vglue -0.5cm
  \caption{Snapshots of the EM energy flux during the inspiral and
    merger of the binary $s_0$. The top and bottom rows show the
    fluxes as measured via~\eqref{FEM_JLmt0} with
    condition~\eqref{first_guess} and via~\eqref{second_guess},
    respectively.  Note that~\eqref{first_guess} yields negative
    values which will cancel out when integrated over one orbit.
    \label{fig:snaps_1}} 
\end{figure*}

While apparently different, expressions~\eqref{first_guess}
and~\eqref{second_guess} lead to very similar estimates and, more
importantly they have a simple interpretation in terms of the
corresponding measures made with. We recall, in fact, that the EM
luminosity is customarily computed via the integral over a 2-sphere of
the Poynting flux $S_i= \sqrt{\gamma}\epsilon_{ijk} E^j B^k$, which is
again just the flux of the stress-energy tensor as measured now by
observers on the spatial hypersurface. Of course also this measure is
adequate only far from the binary and it suffers equally from a
background non-radiative contribution. However, because of the
linearity in the Maxwell equations, the non-radiative contributions
can also be removed by introducing background values of the EM fields
$E_{\rm B}^j, B_{\rm B}^j$ and computing the Poynting vector as $S_i=
\sqrt{\gamma}\epsilon_{ijk} (E^j - E_{\rm B}^j) (B^k - B_{\rm
  B}^k)$. In this context, then, expressions~\eqref{first_guess}
and~\eqref{second_guess} correspond respectively to setting $E_{\rm
  B}^k=0, B_{\rm B}^k=B^k(t=0)$ and $E_{\rm B}^k=E_{\ell, m=0}^k,
B_{\rm B}^k=B_{\ell, m=0}^k$. We have verified that the measures of
the EM luminosity obtained using Equation~\eqref{first_guess}
or~\eqref{second_guess} reproduce well the corresponding ones
obtained with the Poynting flux. As a final remark we note that the EM
flux in Equation~\eqref{FEM_JLmt0} is not always positive everywhere on the
2-sphere. The negative contributions, however, average to zero over
one orbit and do not represent a radiative field. This point was
remarked in ~\cite{Palenzuela:2009hx}, where a toy model within the
membrane paradigm was used for the binary. 

\begin{figure}
  \begin{center}
     \vglue -0.25cm
     \hglue -0.5cm
     \includegraphics[angle=-90,width=4.7cm]{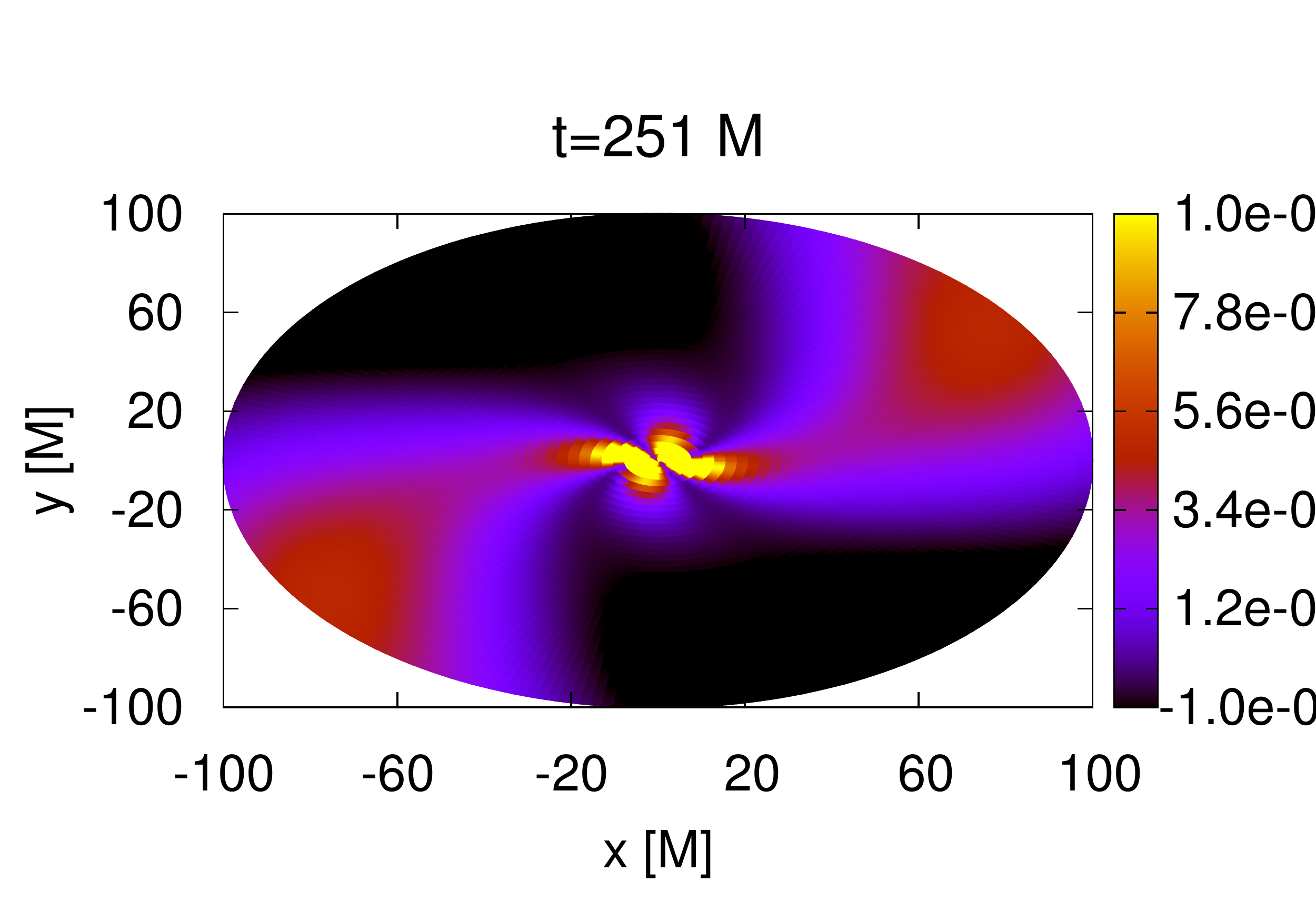}
     \hglue -0.4cm
     \includegraphics[angle=-90,width=4.7cm]{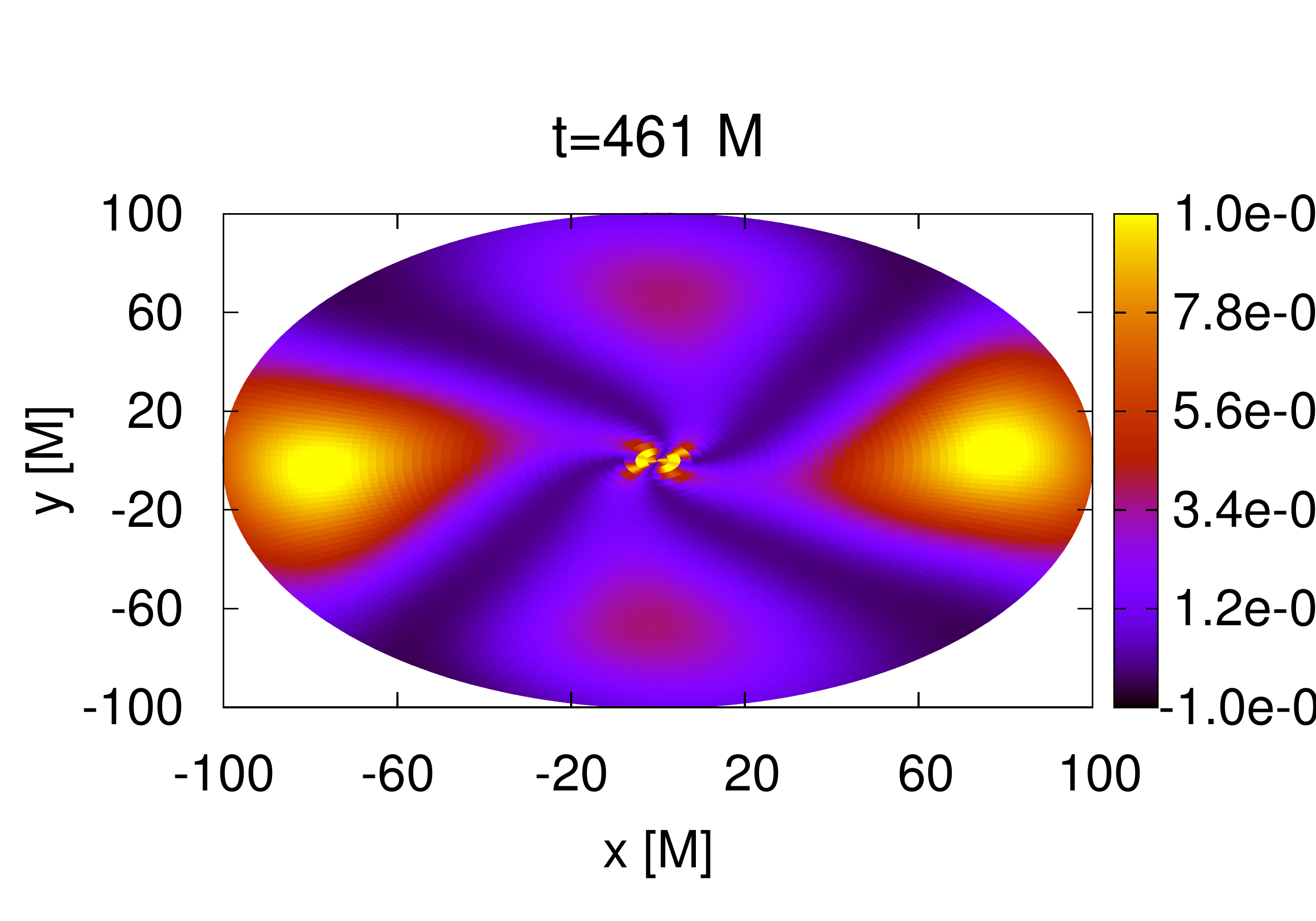}
     \vglue -0.5cm
     \hglue -0.5cm
     \includegraphics[angle=-90,width=4.7cm]{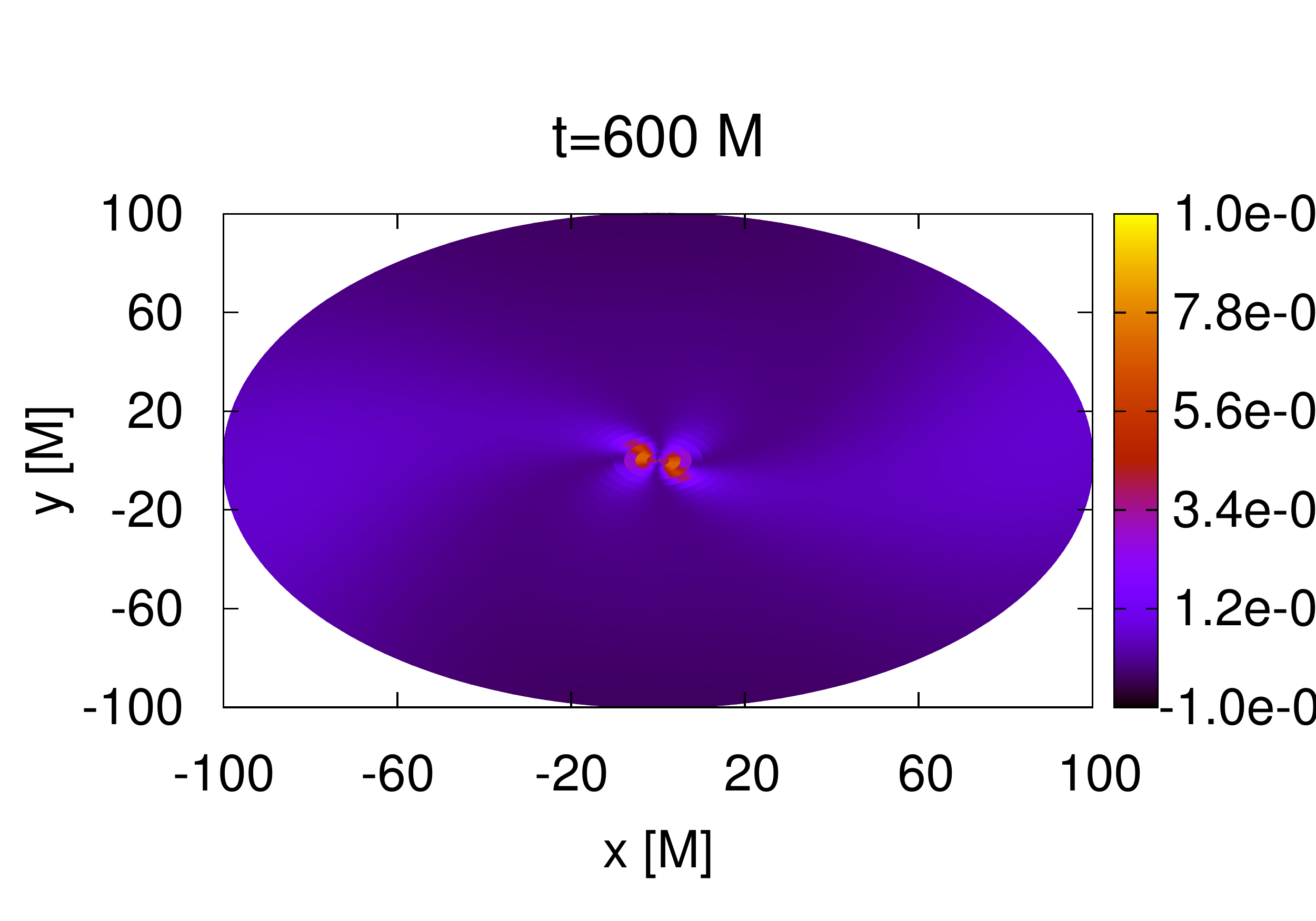}
    \hglue -0.4cm
     \includegraphics[angle=-90,width=4.7cm]{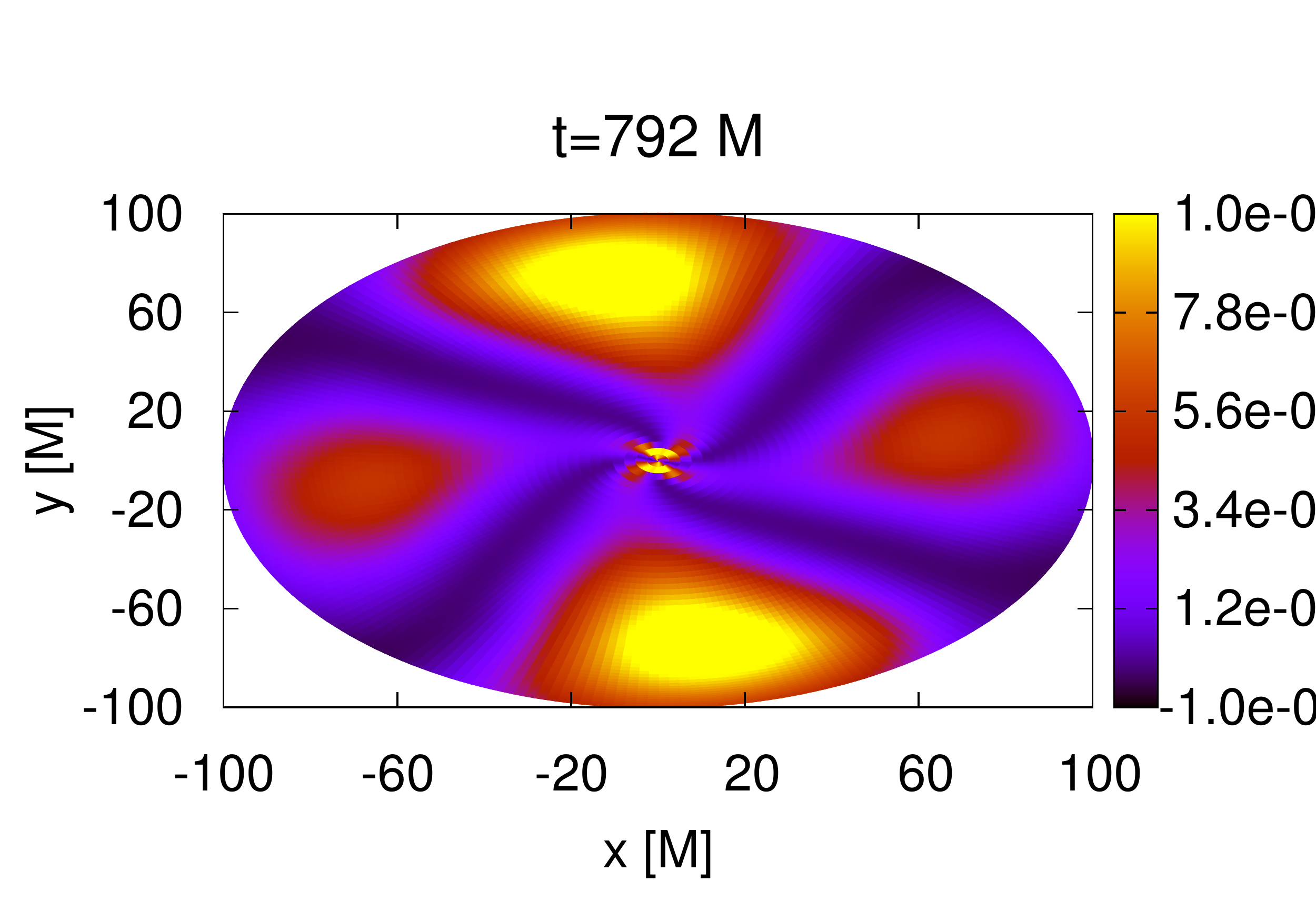}
  \end{center} 
  \vglue -0.5cm
  \caption{Map views of the EM energy flux computed
    using~\eqref{FEM_JLmt0} with~\eqref{first_guess}. The top row
    refers to the binary $s_0$, while the bottom one to the binary
    $s_{6}$ and has a larger emission from the dual jets.
    \label{fig:map}}
  \vglue -0.25cm
\end{figure}

\section{Results}
\begin{figure}
  \vglue -1.5cm
  \begin{center}
     \includegraphics[angle=0,width=8.5cm]{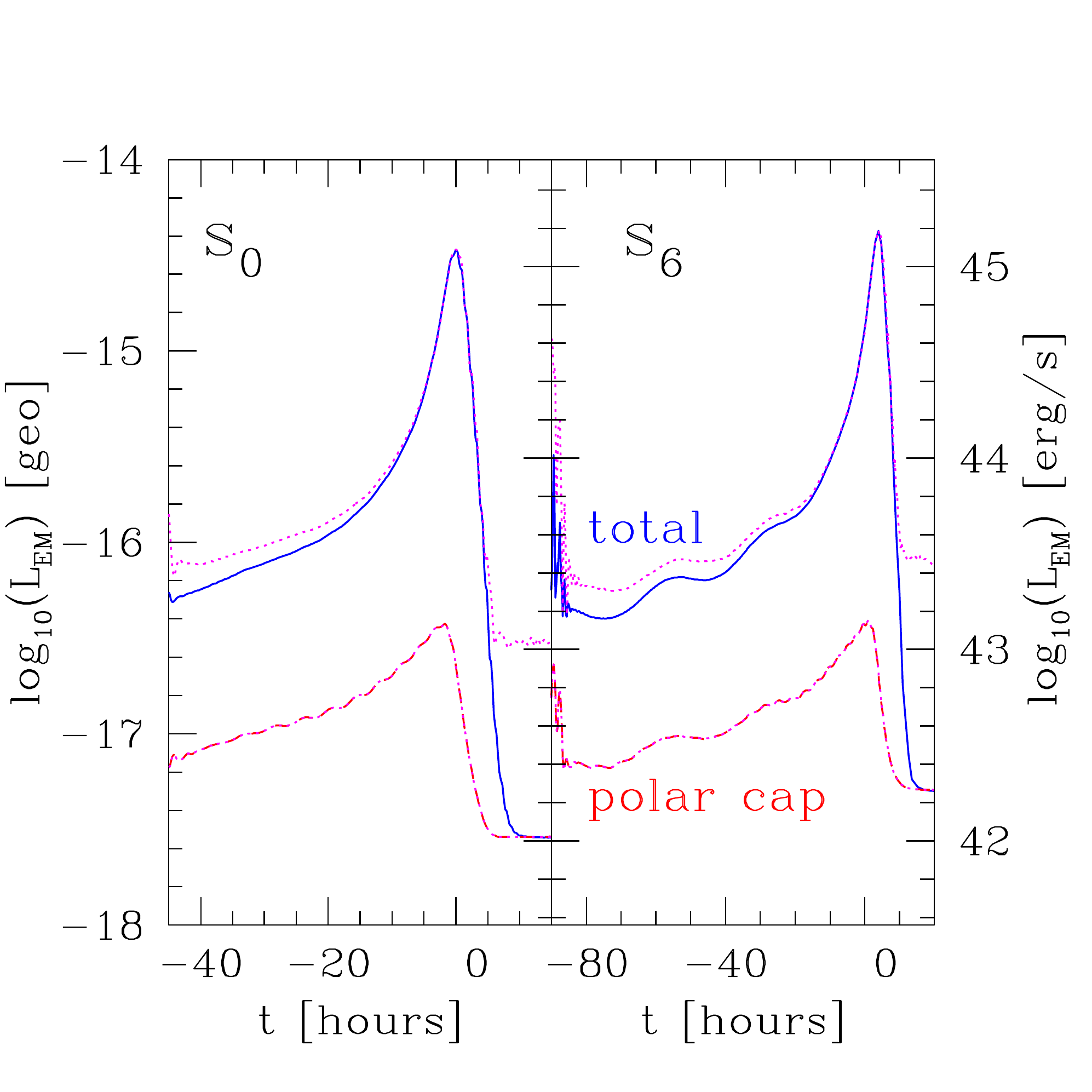}
  \end{center} 
  \vglue -0.5cm
  \caption{Evolution of the EM luminosity at $100\,M$ for the binary
    $s_0$ (left) and for the spinning binary $s_6$ (right), when
    $M=10^8 \, M_{\odot}$ and $B_0=10^4\,{\rm
      G}$. Using~\eqref{second_guess}, (blue) solid lines show the
    total luminosity, while (red) dashed lines refer to the luminosity
    in a polar cap of $5^\circ$. Shown with (magenta) dotted lines are
    the measures with~\eqref{first_guess}; note the presence of a
    small eccentricity for the binary $s_6$.
 \label{fig:Lem}}
\end{figure}

Using expression~\eqref{FEM_JLmt0} with
either the prescription (\ref{first_guess}) or (\ref{second_guess}) we
find that the EM radiation generated during the late inspiral and
merger does contain a dual-jet structure, but also that the total
energy flux is dominated by a non-collimated emission of quadrupolar
nature. This is shown in Figure~\ref{fig:snaps_1}, which reports
snapshots at different times of the EM energy flux on a 2-sphere of
radius $r=100\,M$ for the $s_0$-binary. The top row shows the fluxes
as computed with Equation~\eqref{first_guess} and with a (non-radiative)
component that enhances two lobes of the signal and decreases the
other two (the first, second and third panel refer to the inspiral,
merger and ringdown stages, respectively). Although there are regions
with negative values which will average to zero over one orbit, the
corresponding EM fields could nevertheless induce motion in the plasma
and secondary radiation which we cannot account for here. The second
row refers to fluxes as computed with Equation~\eqref{second_guess}, where the
time dependent background solution is almost completely subtracted, so
that only the (positive) radiative part remains, showing a fairly
symmetric four-lobe structure.  Within both approaches, the extended
quadrupolar distribution is accompanied by the presence of dual jets
during the inspiral, and of a single jet from the spinning merged BH.
The energy flux in the jets is essentially the same in the two cases
(both during the inspiral and after the merger), but that in the
non-collimated part is different. In spite of this, the total
luminosities are very similar, as most of the differences cancel when
integrated over one orbit (\cf Figure~\ref{fig:Lem}). However, because
the Poynting-flux structure is different in the two measures
(Equation~\eqref{first_guess} has a non-radiative part missing
  in Equation~\eqref{second_guess}), it could lead to a different secondary
emission as the EM fields interact with the plasma; this emission
cannot be investigated within an FF approach but deserves attention.

The corrections introduced by the spin of the BHs are shown in
Figure~\ref{fig:map}, which reports map views (\ie projections on the
  $(x,y)$ plane) of the EM emission for the binaries $s_0$ and $s_6$
as computed through Equation~\eqref{second_guess}. Both the collimated and the
non-collimated emission are very similar, although there is a $50\%$
enhancement of the radiation in the spinning case, both in the total
and in the collimated emission. This is because most of the radiation
is produced by the interaction between the BH orbital motion and the
background magnetic field. Indeed we find the emission in the EV
evolution to be comparable to the FF one (this is different from what
reported in~\cite{Palenzuela:2010a, Palenzuela:2010b}).

A more quantitative assessment of the different contributions is shown
in Figure~\ref{fig:Lem}, which reports the evolution of the EM
luminosity of the binary $s_0$ (left panel) and of the spinning binary
$s_6$ (right panel). Shown with (blue) solid lines is the total
luminosity as computed through Equation~\eqref{second_guess}, while the (red)
dashed lines refer to the luminosity integrated over a polar cap with
a half-opening angle of $5^\circ$ and thus representative of the
emission from the two jets. Also shown with (magenta) dotted lines are
the luminosities as computed through Equation~\eqref{first_guess}: the
differences are small and hardly visible for the collimated part.

An accurate measure of the evolution of the collimated/non-collimated
components is crucial to predict the properties of the system when the
two BHs are widely separated. This, in turn, requires a reliable
disentanglement of the collimated emission from the non-collimated one
and from the background. The non-collimated emission measured
with Equation~\eqref{second_guess} matches well the growth expected if the EM
emission is mostly quadrupolar and hence with a dependence that is the
same as the GW one and, at the lowest order, scales as $L^{\rm
  non-coll}_{_{\rm EM}} \approx
\Omega^{10/3}$~\cite{Palenzuela:2009hx, Moesta:2009}. On the other
hand, a smaller scaling is found when the approach Equation~\eqref{first_guess}
is adopted; at present it is difficult to determine which one is the 
most reliable scaling. At the same time, the frequency evolution for the
collimated emission coincides in the two approaches, but it does not
follow the scaling $L^{\rm coll}_{_{\rm EM}} \approx \Omega^{2/3}$
suggested in~\cite{Palenzuela:2010a} for boosted BHs; rather, it
scales as $L^{\rm coll}_{_{\rm EM}} \lesssim \Omega^{5/3-6/3}$. We
suspect the accelerated motion of the BHs to be behind this difference
and longer simulations are needed to draw robust conclusions.

\noindent\emph{Impact on detectability.~}Assessing the detectability
of the EM emission discussed above is of course of great importance
and using the results of~\cite{Palenzuela:2010a, Palenzuela:2010b},
~\cite{Kaplan:2011} have estimated that the short timescales
associated with the merger will limit the detectability in the radio
band to less than $1 \rm{yr}^{-1}$. As reported in Figure~\ref{fig:Lem}, the
peak of the collimated emission is $\sim 100$ times smaller than that
of the total emission, making the detection of the dual jets at the
merger unlikely. Unfortunately, even if the energy flux is $\sim 8-2$
times larger near the jets, the lack of knowledge about the Lorentz
factor of the reprocessed plasma does not allow us to say whether the
beaming in the jet will be larger than that in the extended emission
and thus help its detection. That said, because the total luminosity
at merger is $\sim 100$ times larger than in~\cite{Palenzuela:2010a}
(mainly because we measure it at large distances where it has
approximately reached its asymptotic value), the detection should be
overall more likely if most of the assumptions in~\cite{Kaplan:2011}
are verified. In addition, the dual jets emission could be dominant in
the early inspiral (especially if the BHs are spinning). Assuming the
scalings for the collimated and non-collimated emissions are different
and go respectively like $\Omega^{10/3}$ and $\Omega^{6/3}$
$(\Omega^{5/3})$, the two components of the luminosity would become of
the same order at a separation of $\sim 16~(24)\,M$, but obviously
smaller (the collimated will be smaller by a factor $\sim
10$). Determining more precisely when and if this happens requires an
accurate frequency scaling which is not available yet.  As a note of
caution we stress that luminosities $L_{_{\rm EM}}\sim 10^{45}\,{\rm
  erg\ s^{-1}}$ are also typical of radio-loud galaxies, and the
determination of an EM counterpart can be challenging if such sources
are near the candidate event.

\section{Conclusions}We have investigated the suggestion that
dual jets can be produced during the inspiral and merger of
supermassive BHs immersed in an FF plasma threaded by a uniform
magnetic field. We have found that the energy flux does have a
dual-jet structure but is predominantly quadrupolar, with the
non-collimated emission being about $10-100$ times larger than the
collimated one. Our findings set restrictions on the detectability of
dual jets from coalescing BH binaries, but also increase the chances
of detecting an EM counterpart for astrophysical conditions similar to
those in this simplified scenario.

\medskip
\section{Acknowledgments}We are grateful to J.-L. Jaramillo
for discussions and to L. Lehner for suggesting an approach similar
to Equation~\eqref{first_guess}.  We also thank A.  Sesana, B. Ahmedov, and
V. Morozova. This work was supported in part by the DFG grant
SFB/Transregio~7; the computations were made at the AEI and on
TERAGRID (TG-MCA02N014).



\end{document}